\documentclass[pra,twocolumn,showpacs,superscriptaddress]{revtex4}
\pdfoutput=1
\usepackage[english]{babel}
\usepackage{graphicx}
\usepackage{amsmath,amssymb}
\usepackage[utf8]{inputenc}
\usepackage{natbib}
\usepackage{units}
\usepackage{hyperref}
\usepackage{color}
\usepackage{url}
\hypersetup{colorlinks=true,citecolor={blue},linkcolor={blue},urlcolor={blue}}
\usepackage{notes2bib}
\usepackage{bm}
\def\cc{{\rm cav}}
\def\dd{{\rm def}}

\def\vect#1{\bm{#1}}

\newif\iffig\figfalse
\begin{document}
\title{Linear Wave Dynamics Explains Observations Attributed to Dark Solitons \\in a Polariton Quantum Fluid}

\date{\today}

\author{P. Cilibrizzi}
\affiliation{School of Physics and Astronomy, University of Southampton, Southampton, SO17
 1BJ, United Kingdom}

\author{H. Ohadi}
\affiliation{School of Physics and Astronomy, University of Southampton, Southampton, SO17
 1BJ, United Kingdom}

\author{T. Ostatnicky}
\affiliation{Faculty of Mathematics and Physics, Charles University in Prague, Ke Karlovu 3, 121 15 Prague, Czech Republic}

\author{A. Askitopoulos}
\affiliation{School of Physics and Astronomy, University of Southampton, Southampton, SO17
 1BJ, United Kingdom}

\author{W. Langbein}
\affiliation{School of Physics and Astronomy, Cardiff University, The Parade, CF24 3AA Cardiff, United Kingdom}

\author{P. Lagoudakis}
\email[correspondence address:~]{pavlos.lagoudakis@soton.ac.uk}
\affiliation{School of Physics and Astronomy, University of Southampton, Southampton, SO17
 1BJ, United Kingdom}

\begin{abstract}
We investigate the propagation and scattering of polaritons in a planar GaAs microcavity in the linear regime under resonant excitation. The propagation of the coherent polariton wave across an extended defect creates phase and intensity patterns with identical qualitative features previously attributed to dark and half-dark solitons of polaritons. We demonstrate that these features are observed for negligible nonlinearity (i.e., polariton-polariton interaction) and are, therefore, not sufficient to identify dark and half-dark solitons. A linear model based on the Maxwell equations is shown to reproduce the experimental observations.
\end{abstract}

\maketitle

Solitons are solitary waves that preserve their shape while propagating through a dispersive medium~\cite{stegeman_optical_1999,segev_self-trapping_1998} due to the compensation of the dispersion-induced broadening by the nonlinearity of the medium~\cite{chen_optical_2012}. Over the years, spatial solitons have been observed by employing a variety of nonlinearities ranging from Kerr nonlinear media~\cite{Barthelemy1985201} to photorefractive~\cite{segev_spatial_1992} and quadratic~\cite{hayata_multidimensional_1993} materials. Apart from their potential application in optical communications~\cite{Nakazawa_392661,PhysRevE_53_4137}, solitons are important features of interacting Bose-Einstein condensates (BECs) and superfluids. The nonlinear properties of BECs can give rise to the formation of quantized interacting vortices and solitons, the latter resulting from the cancellation of the dispersion by interactions, for example, in atomic condensates.
A special class of solitons is the so-called dark soliton, which feature a density node accompanied by a $\pi$ phase jump. Since the first theoretical prediction in the context of BECs~\cite{tsuzuki_nonlinear_1971}, dark solitons were studied and observed first in the field of nonlinear optics~\cite{kivshar_dark_1993} and, then, in cold-atom BECs~\cite{denschlag_generating_2000}.
The experimental observation of BECs~\cite{kasprzak_boseeinstein_2006} and superfluidity~\cite{amo_superfluidity_2009,amo_collective_2009} of exciton-polaritons, has sparked interest in the quantum-hydrodynamic properties of polariton fluids. In particular, the nucleation of solitary waves in the wake of an obstacle (i.e.~defect) has been claimed recently~\cite{amo_polariton_2011,PhysRevLett.107.245301,PhysRevB.86.020509,deveau_book,hivet_half-solitons_2012}. Here, the source of nonlinearity, essential for the formation of such a solitary wave, has been identified in the repulsive polariton-polariton interactions.
\begin{figure}[!hbtp]
\includegraphics[scale=0.45]{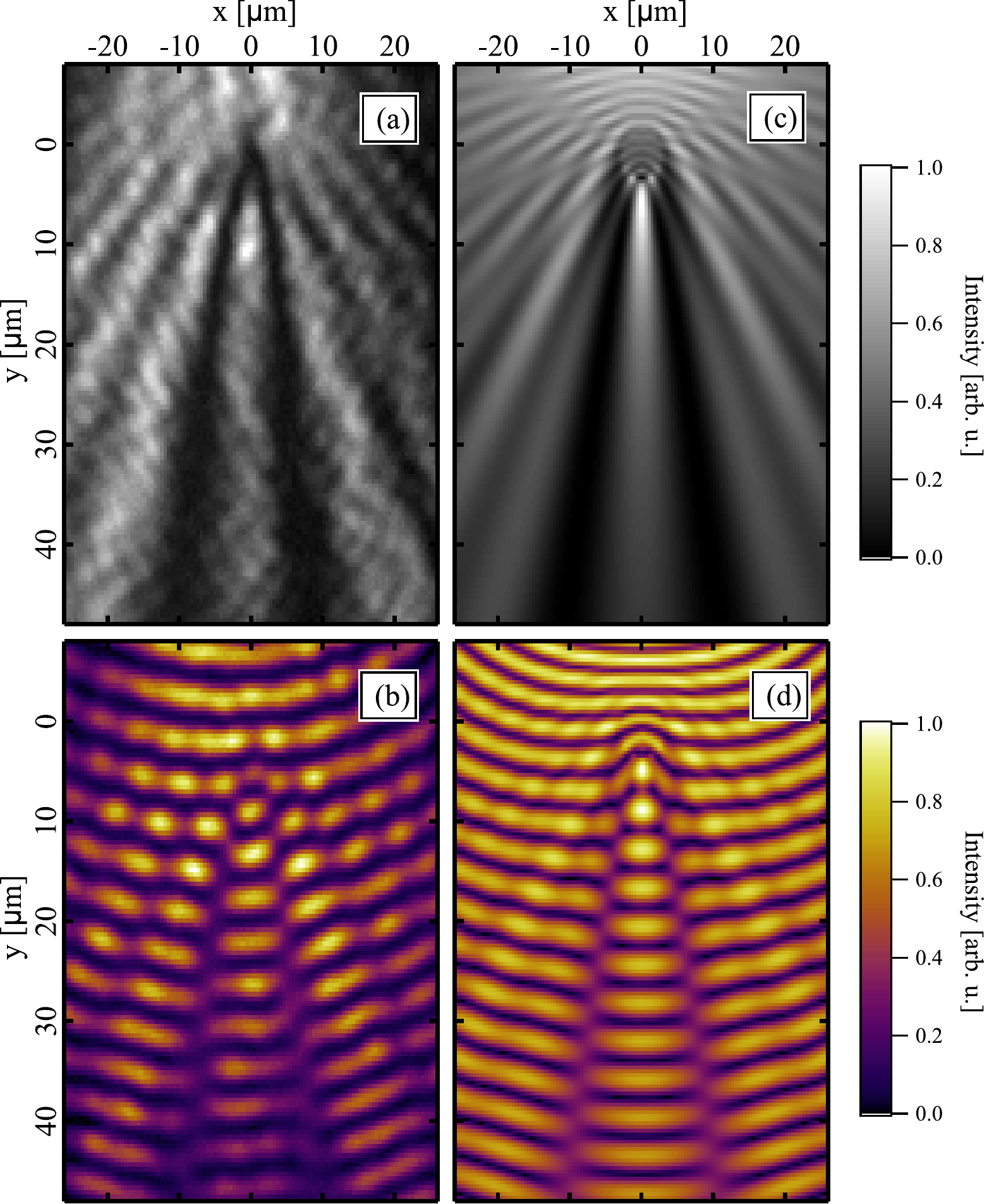}
\caption{Experimental (a),(b) and simulated (c),(d) real space intensity and interference patterns showing the two "soliton fingerprints" generated by the scattering of a beam with a pointlike defect: a dark notch in the intensity pattern together with $\pi$ phase dislocations. In the images, the polaritons propagating downwards, along the \textit{y} axis, are injected with a wave vector of $\unit[1.5] {\mu m}^{-1}$ and are scattered by a defect positioned $\unit[25] {\mu m}$ away from the excitation spots.}
\label{fig:1}
\end{figure}
In these previous works, the observation of dark notches in the intensity profiles together with a $\pi$ shift in the phase have been used as sufficient signatures for dark solitons in microcavities. In addition, half-dark solitons have been reported to carry a nonzero degree of circular polarization in the presence of the TE-TM splitting of the cavity mode~\cite{hivet_half-solitons_2012}.
\newline
In this Letter, we demonstrate that these features, used as dark-soliton fingerprints~\cite{amo_polariton_2011,PhysRevLett.107.245301, PhysRevB.86.020509,deveau_book,hivet_half-solitons_2012}, can also be observed without the presence of nonlinearity, which is the fundamental ingredient differentiating solitons from linear wave propagation. Specifically, we investigate the propagation of polaritons with a small exciton fraction and at low polariton densities, excluding a relevant influence of nonlinearities.
We show that polariton propagation in this linear regime across an extended defect can create deep notches in the intensity profile accompanied by a $\pi$ phase shift.
We model the observation using linear wave propagation, clarifying that these features are not indicative of a nonlinear interaction between polaritons, but are interference patterns created by scattering from the defect. Moreover, we show that the appearance or disappearance of these features for different in-plane kinetic energies is found also in the linear regime and, thus, does not provide evidence of an interacting quantum fluid. Therefore, the previous reports of the observation of dark solitons~\cite{amo_polariton_2011,PhysRevLett.107.245301,PhysRevB.86.020509,deveau_book} and half-dark-solitons~\cite{hivet_half-solitons_2012}, which were based on these features, have to be reconsidered.
\newline The investigated sample is a bulk $\lambda$ GaAs microcavity surrounded by 27 (top) and 24 (bottom) distributed GaAs/AlAs Bragg reflector pairs. The sample is held in a cold-finger cryostat at a temperature of 15\,K and is illuminated by a narrow linewidth single-mode continuous wave laser, tuned to the resonance of the cavity at about $\unit[1.485]{eV}$. The measurements were performed in transmission configuration. The phase was measured using a shearing Mach-Zehnder interferometer (see~\cite{suppl_info}, S1). Our experiments were performed in the linear regime, facilitated by the large negative detuning of $\unit[-29]{meV}$ of the cavity photon mode from the exciton resonance at \unit[1.514]{eV}, resulting in a small exciton fraction of the polariton of about 1\%.
To verify the linear regime, we studied the excitation density dependence of our results with both a Gaussian and half-Gaussian excitation beam (see~\cite{suppl_info}, S2). We find that they are independent of both the shape of the beam and the excitation density over a range of 4 orders of magnitude and they persist at polariton density as low as $\unit[2.3\times 10^2]{cm^{-2}}$, 7 orders of magnitude lower than the lasing threshold observed in standard microcavities~\cite{Deng23122003}. 
\newline The real space intensity and interference of a polariton wave propagating across a defect are shown in Fig.~\ref{fig:1}. The experimental results show the presence of two dark notches in the intensity pattern along with a $\pi$ phase shift visible in Fig.~\ref{fig:1}(b) as paths of vortices merging in succession with alternating topological charge $\pm 1$.
\begin{figure}[!hbtp]
\includegraphics[scale=0.25]{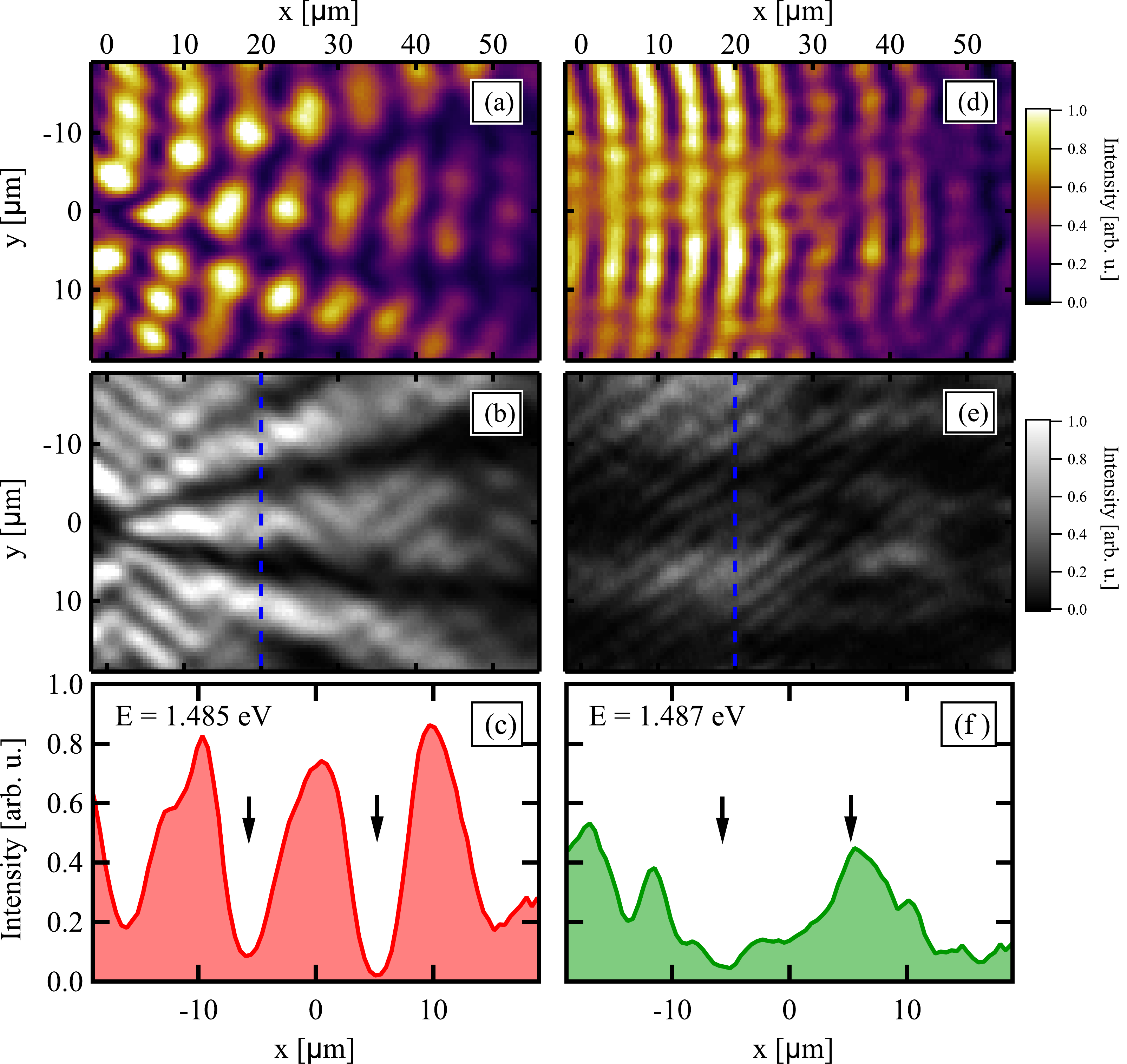}
\caption{Experimental interference (a),(c) and intensity (b),(d) showing the transition between the regime where the soliton features are well defined ($\unit[1.485]{eV}$) to a regime where they vanish ($\unit[1.487]{eV}$). The intensity profiles (e),(f) calculated along the blue dashed line, $\unit[20] {\mu m}$ away from the defect, confirm that the dark notches disappear when the energy of the excitation beam is increased. The two arrows indicate the positions of solitonlike fingerprints.}
\label{fig:2}
\end{figure}
Simulations of the measurements using the realistic experimental parameters are shown in Figs.~\ref{fig:1}(c) and \ref{fig:1}(d). 

Solitons are predicted to appear in polariton microcavities as the result of the nonlinearity due to the polariton-polariton interactions~\cite{pigeon_hydrodynamic_2011}. Since our experiments are in the linear regime, it is important to understand how the nature and the size of the defect affects the formation process of these solitonlike features.  In a recent study~\cite{zajac_polariton_2012} of the structural and optical properties of GaAs/AlAs microcavities grown by molecular beam epitaxy it was shown that the most common pointlike defects were characterized by a circular or elliptical shape~\cite{zajac_structure_2012}, due to Gallium droplets emitted occasionally during the growth~\cite{fujiwara_classification_1987,chand_comprehensive_1990}. The presence of the defect has the effect of modifying the effective thickness of the cavity layer, which typically results in an attractive potential for the cavity mode inside the defect~\cite{zajac_structure_2012}. Consequently, the wave vector of the photonic mode in the region of the defect is higher than in the rest of the cavity.

The polariton scattering by the defect depends on the wave vector mismatch between the polaritons outside and inside the defect at the energy of excitation.  When the energy shift of the defect photon mode with respect to the unperturbed cavity mode is large enough to make the coupling between them inefficient, the defect behaves like a hard scatterer and the spatial intensity distribution is similar to the complementary case of a single-slit diffraction~\cite{hecht_optics_1982}. In our case, however, there is a finite transmission through the defect, producing dark and bright traces with a more complicated phase pattern. 
As has been shown by Berry~\textit{et al.}~\cite{Berry_rays_1992, berry_elliptic_1979}, wave fronts resulting from interference can contain dislocation lines. In the case of a scattered beam, dislocations are composed of phase shifts at positions where the amplitude of the electromagnetic wave and, thus, the intensity vanishes, representing nodes of the wave. It is worth mentioning that nonlinearities are negligible close to nodes also in the nonlinear regime, and phase dislocations at zero intensity (i.e.,~at the dark notches) are features of both linear~\cite{p._senthilkumaran_two_2010,ruben_phase_2007} and nonlinear waves. In our case, the analogy between linear and nonlinear waves goes beyond the mere observation of the same features and is effectively more profound. Indeed, as shown in~\cite{suppl_info} S$4$, the intensity, the phase jumps as well as the relative depth of the dark notches in the linear regime satisfy the same mathematical expression as in the quantum fluid case~\cite{amo_polariton_2011,PhysRevLett.107.245301,PhysRevB.86.020509,deveau_book}. In particular, also in our linear system the relative depth of the dark notches remains constant up to $42~\mu$m (see~\cite{suppl_info}, S4).  

Beyond the qualitative discussion above, we performed simulations of the experiments, based on a numerical solution of the linear scattering problem using the classical theory of electromagnetism. The choice of such a model is justified by the fact that we operate in the linear regime and with a small exciton fraction of about 1\%, such that the polariton dispersion is dominated by the cavity mode. In the model, we consider the propagation of quasi-two-dimensional photons with a parabolic dispersion in a cavity with a fixed width. The incident wave has been treated as coming from a linearly polarized pointlike source with polarization in the plane of the cavity. Defects have been modeled as disk-shaped perturbations of the cavity thickness resulting in an energy shift of the photon dispersion (see~\cite{suppl_info}, S5). To model the defect parameters, which are not experimentally known, we use a disk shape with a radius of $\unit[3]{~\mu m}$ and a polariton potential of $-2.3$ meV (consistent with Ref.~\cite{zajac_structure_2012}). Maxwell's equations are then solved using an expansion of the fields into the planar cavity eigenmodes in cylindrical coordinates fulfilling the boundary conditions for tangent components of electric and magnetic fields on the interface between the cavity and the defect (see~\cite{suppl_info}, S5). This linear wave dynamics model reproduces the intensity notch and the phase dislocation previously used as dark-soliton fingerprints. The results show a marked dependence on the geometry of the scattering problem, as shown in S6~\cite{suppl_info}.  In particular, the phase jump visible in the interference pattern depends on the direction of the incoming polariton wave relative to the defect (see Fig.\,S7). On the other hand, the size of the defect relative to the polariton wavelength affects the formation of high-order phase dislocations (see Fig.\,S8).
%
\begin{figure}[!hbtp]
\includegraphics[scale=0.48]{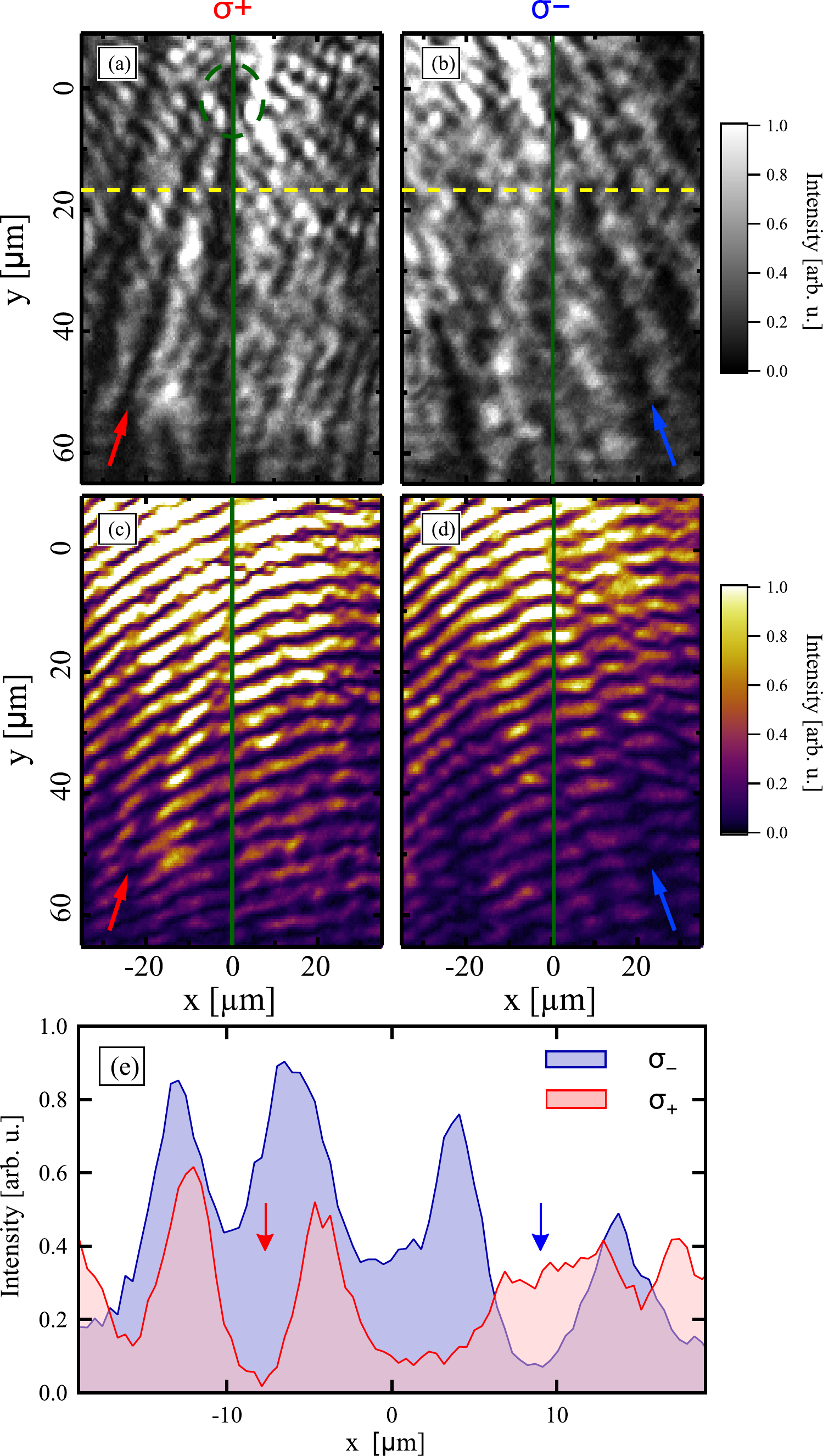}
\caption{Experimental intensity pattern (a)-(b) and real space interference (c)-(d) showing two half-soliton features as indicated by the arrows. The red and blue arrows indicate, respectively, the position of the $\sigma_+$ and $\sigma_-$ soliton features: a dark notch with an associated phase jump present in only one circular component. The green vertical line is a guide for the eyes to distinguish the two different regions while the dashed circle in (a) indicates the defect. (e) The intensity profile extracts from the yellow dotted line displaying the two dark notches present, respectively, in only one opposite polarization basis, as indicated by the arrows.}
\label{fig:3}
\end{figure}
\\
In a nonlinear cavity-polariton system, a polariton fluid has been predicted to flow almost unperturbed around the defect (i.e.~disappearance of the features) or experience the nucleation of vortices and/or solitons at the position of the defect (i.e.~appearance of the features), depending on the excitation density or on the energy of the pump~\cite{pigeon_hydrodynamic_2011}. We evaluated the possibility of observing these features, ascribed in the literature to dark solitons resulting from the interaction within the polariton fluid, in the absence of non-nonlinearities. Figures~\ref{fig:2}(a) and \ref{fig:2}(b) show the phase and the intensity of solitonlike fingerprints in real space. Instead of increasing the excitation power, which has no effect in the linear regime, we tune the energy of the excitation beam and observe the appearance and disappearance of solitonlike features. As discussed above, the appearance of the intensity minima and phase dislocations is a result of interference which is sensitive to the intensity and relative phase of contributing waves. The increase of the energy of the excitation beam by 2~meV causes an increase of the in-plane wave vector of the propagating polariton mode that, in turn, changes the interference condition so that the straight dark notches  [Fig.~\ref{fig:2}(c)] and the phase dislocations [Fig.~\ref{fig:2}(d)] disappear. The wave vector dependence of such transitions will depend on the defect structure and the related bound polariton states~\cite{zajac_structure_2012}, so that they could also be observed with decreasing wave vector for other defects. 
Intensity profiles measured at a fixed distance from the defect [Figs.~\ref{fig:2}(e) and \ref{fig:2}(f)] confirm the observed transition. Thus, it becomes apparent that the appearance or disappearance of solitonlike features, although independent of the excitation density, strongly depends on the wave vector of the propagating mode (see~\cite{suppl_info}, S3). It is worth noting that an increase of the polariton density corresponds to an energy blueshift of the polariton dispersion. For polaritons excited resonantly with a given energy, this results in an increase of the polariton wave vector with decreasing density along the polariton propagation. Specifically, in nonresonantly excited experiments~\cite{NatPhys_tosi_2012}, this blueshift is dominated by the exciton density in the reservoir at high wave vectors. The interaction with the exciton reservoir is not a polariton-polariton interaction within the condensate which could provide the nonlinearity needed for the formation of solitons but, instead, represents an external potential sculpting the polariton energy and gain landscape.

In a different experiment, we address the observation of half-soliton fingerprints, which requires polarization-resolved measurements. The intensity images [Figs.~\ref{fig:3}(a) and \ref{fig:3}(b)] are measured using an excitation linearly polarized parallel to the \textit{y} direction. The interferograms [Figs.~\ref{fig:3}(c) and \ref{fig:3}(d)] are obtained by selecting the same polarization for the excitation and reference beam (see~\cite{suppl_info}, S1 for details). The signature of an oblique dark half-soliton is a notch in only one circular polarization component~\cite{flayac_oblique_2011, hivet_half-solitons_2012}. We excite the sample with a linearly polarized beam and detect the two circular polarization components ($\sigma_-$, $\sigma_+$) separately. The measurements are performed with the same excitation energy ($\unit[1.485]{eV}$) and negative detuning ($\unit[-29]{meV}$) as in the previous case.
The measured intensity and the interferogram for the $\sigma_-$ component are given in Figs.~\ref{fig:3}(a) and \ref{fig:3}(c), respectively. The images show the presence of a $\sigma_-$ soliton fingerprint, indicated by the blue arrows, that is absent in the $\sigma_+$ component [Figs.~\ref{fig:3}(b) and \ref{fig:3}(d)]. The same applies to the $\sigma_+$ counterpart, where a half-soliton fingerprint is observed only on the right side of the image.

\begin{figure}[!hbtp]
\includegraphics[scale=0.45]{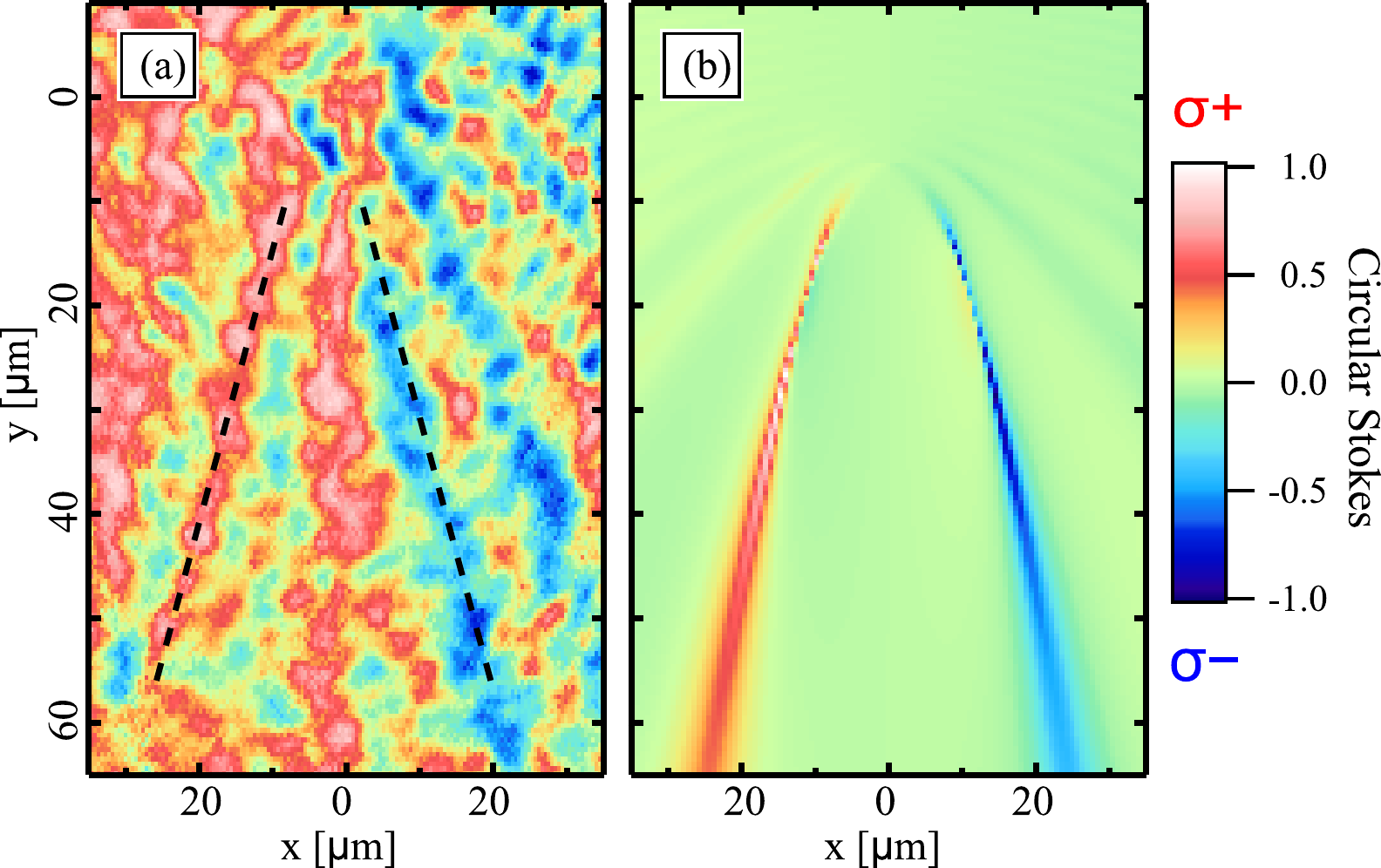}
\caption{Experimental (a) and simulated (b) circular Stokes parameter showing half-soliton features. The two black dotted lines correspond to the position of the dark notches present in Figs.~\ref{fig:3}(a) and \ref{fig:3}(b).}
\label{fig:4}
\end{figure}

By calculating the degree of circular polarization, given by $S_c=(I_{\sigma_+} - I_{\sigma_-})/(I_{\sigma_+} + I_{\sigma_-})$, with $I_{\sigma_+}$ and $I_{\sigma_-}$ being the measured intensities of the two components, we measure the pseudospin state inside the cavity [Fig.~\ref{fig:4}]. Here, if we look at the same position where the soliton features have been observed [Fig.~\ref{fig:3}] indicated by the black dotted lines in Fig.~\ref{fig:4}(a), we note the presence of a pair of oblique traces with opposite circular polarization, resembling the predictions and observations attributed to a polariton superfluid~\cite{flayac_oblique_2011, hivet_half-solitons_2012}. The high degree of circular polarization that we observe is due to the polarization splitting of transverse electric and transverse magnetic optical modes (TE-TM splitting)~\cite{panzarini_exciton-light_1999} (see~\cite{suppl_info}, S7). The latter gives rise to the optical spin Hall effect~\cite{kavokin_optical_2005} that has been observed in both polaritonic~\cite{leyder_observation_2007} and photonic microcavities~\cite{maragkou_optical_2011}.
In our simulations [Fig.~\ref{fig:4}(b)], a linearly polarized incoming beam propagates along the \textit{y} direction and is scattered by a defect positioned at $\unit[25]{\mu m}$ away from the excitation spot, inducing the formation of two traces propagating in oblique directions.
The detected field is a superposition of the incoming linearly polarized wave and the scattered wave. The TE-TM splitting of the optical mode in a photonic cavity is responsible for an anisotropy in the polarization flux, as previously shown on the same sample in Ref.~\cite{maragkou_optical_2011}. Here, the same values of the TE-TM splitting have been used to perform the simulations. 
The polaritons scatter from the defect with wave vectors of equal modulus but in different directions both in the real and momentum space. Because of the birefringence induced by the TE-TM splitting, polaritons propagating in different directions experience different polarization rotation and shift. Polaritons traveling to the right gain a $\sigma_+$ component while polaritons traveling to the left gain a $\sigma_-$ component. The anisotropy of the effect manifests itself in the intensity pattern, where it is possible to observe the features of an oblique soliton in one circular component and not in the other.

In conclusion, we have shown that the previously reported experimental signatures of oblique dark solitons and half-solitons in polariton condensates can be observed in the case of polaritons propagating in the linear regime. In our experiments these features are the result of the interference of the incoming wave with the waves scattered by the defect. Their phase jumps and the relative depth of the dark notches satisfy the same analytical expression as in the polariton quantum fluid. In the case of the polarized counterpart (i.e., half-solitonlike features) the intrinsic TE-TM splitting of the cavity dispersion gives rise to oblique straight traces with opposite polarization.

Our results clarify that phase vortex lines in polariton propagation together with dark notches of constant relative depth in the intensity patterns, used as fingerprints of oblique-dark solitons and half-solitons in the literature, are present in the linear propagation regime. Consequently, these features are necessary, but not sufficient, evidence to identify solitons.
We believe a more reliable criterion for identifying dark solitons, based on the definition of solitons (i.e., solitary nonspreading wave), would be the size of the observed features which should be determined by the healing length of the condensate (see~\cite{suppl_info}, S$4$ for details). 
\\
\\
P.C. and P.L. acknowledge the Marie Curie ITNs Spinoptronics for funding. A.A. and P.L. acknowledge funding from Marie Curie ITN Clermont IV 235114. T.O. acknowledges financial support from the Grant Agency of Czech Republic, Project No. P204/10/P326. W.L. and P.L. acknowledge support by the EPSRC under Grant No. EP/F027958/1. P.C. acknowledges stimulating discussions with Stefano Portolan.
\setcounter{figure}{0} \renewcommand{\thefigure}{\textbf{S\arabic{figure}}}
\renewcommand{\thesection}{S\arabic{section}}

\cleardoublepage
\begin{widetext}

\begin{center}
{\large\textbf{Supplementary Information}}
\end{center}

\hspace{1.0cm}

\section{Experimental setup}
\label{sec:1}

The investigated sample is a $\lambda$~bulk GaAs microcavity surrounded by 27 (top) and 24 (bottom) distributed GaAs/AlAs Bragg reflector pairs (DBRs).
The sample is held in a cold-finger cryostat at a temperature of 15\,K and is excited by a narrow linewidth ($\approx~30~\unit{kHz}$) single-mode continuous wave Ti:Sapphire laser, tuned to the cavity resonance at about $\unit[1.485] {eV}$ depending on the excitation in-plane wavevector.
\begin{figure}[h]
\includegraphics[scale=0.4]{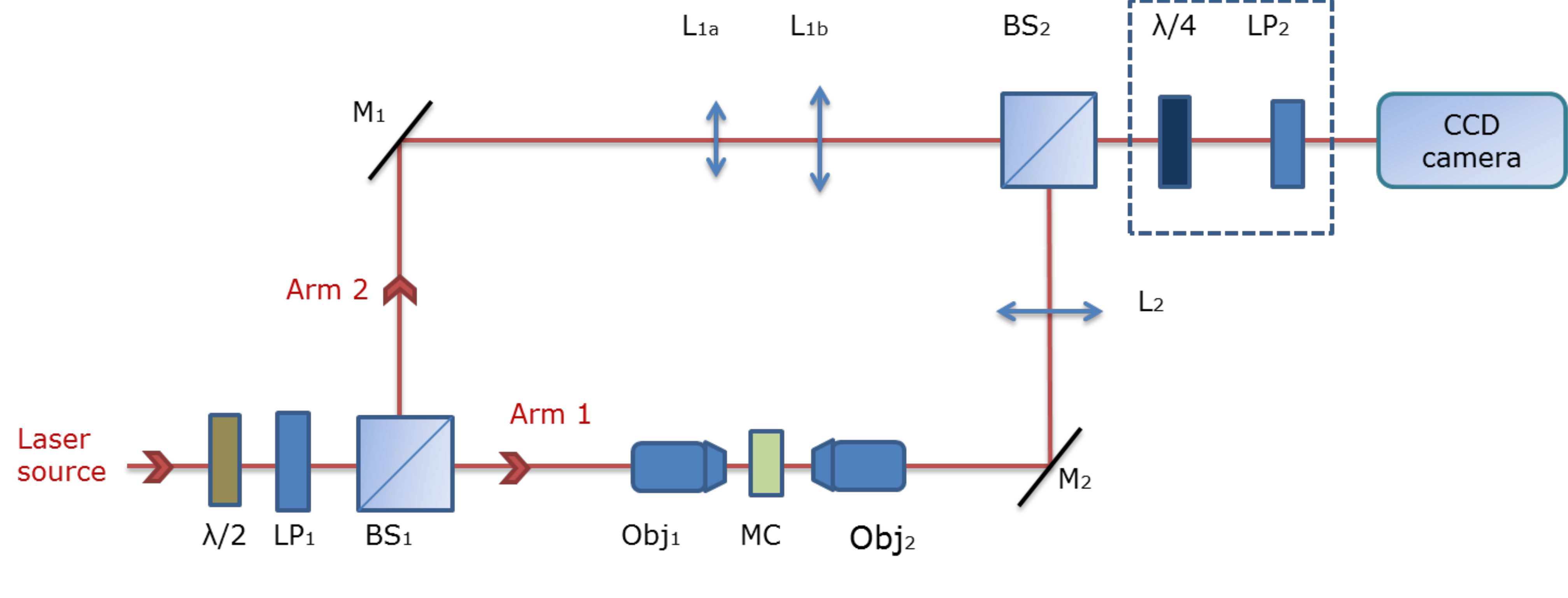}
\caption{Sketch of the Mach-Zehnder interferometer used in the experiments. List of the optical components: BS$_1$ and BS$_2$ are non-polarizing beam splitters; M$_1$ and M$_2$ are mirrors; Obj$_1$ is the excitation objective with a 20x magnification and 0.4 numerical aperture; Obj$_2$ is the objective used for collection of the transmitted beam, with 10x magnification and 0.25 numerical aperture; L$_{1a}$, L$_{1b}$ and L$_2$ are convex lenses;  $\lambda/4$ and LP$_2$ are respectively the quarter-wave plate and the linear polarizer used to measure the circular Stokes parameters while $\lambda/2$ and LP$_1$ are respectively a half-wave plate and a linear polarizer used to control the excitation power. The optical elements enclosed in the dashed box ($\lambda/4$ and LP$_2$) are introduced in the setup only for polarization resolved measurements used to obtain the data shown in Figs.~3(a-d) of the manuscript.}
\label{fig:s1}
\end{figure}

The data reported in the manuscript have been acquired by using the experimental setup represented in Fig.~\ref{fig:s1}. The intensity measurements have been performed in transmission configuration, by blocking Arm 2 of the interferometer. The interferograms revealing the phase, on the other hand, are acquired using both Arm 1 and Arm 2. In this arrangement the setup corresponds to a Mach-Zehnder interferometer, in which a laser is split into two arms: one is used to excite the sample (Arm 1), while the other with a flat phase is used as the reference (Arm 2).
The excitation beam (Arm 1) is focused by a 0.4 NA microscope objective to a spot on the sample with a full width at half maximum (FWHM) $2\sqrt{\ln2}\sigma$ of $3~\unit{\mu m}$, resulting in a circular distribution in momentum space (i.e. \textit{k}-space) with a diameter of $ 3~\unit{\mu m^{-1}}$. The light transmitted through the sample is then collected using a 0.25 NA microscope objective and focused on a charge-coupled device (CCD) camera by a convex lens ($L_2$). In this way, the transmitted beam is imaged in real space.  The reference beam (Arm 2) is expanded by a telescope (formed by L$_1a$ and L$_1b$ in Fig.~\ref{fig:s1}) so that a bigger area of the sample could be investigated and then interfered with the transmitted beam on the CCD camera. The incidence angle of the reference beam was adjusted in order to obtain interference fringes along $y$. The power of the excitation beam is adjusted by means of a half-wave plate ($\lambda/2$) and a linear polarizer (LP$_1$).
\\
\\
\textbf{Polarization resolved measurements}. The investigation of half-soliton features, shown in Figs.~3(a-d) of the manuscript, requires polarization-resolved measurements. In our setup, the linear polarizer (LP$_1$) prepares the excitation beam in the linearly polarized basis (parallel to the \textit{y} axis in Figs.~3 of the manuscript) and we introduce in the setup a polarimeter composed of a quarter-wave plate ($\lambda/4$) and a linear-polarizer (LP$_2$), oriented at $45^{\circ}$ with respect to one another, to measure the circular Stokes parameter of the transmitted signal. In this way, by rotating the wave-plate it is possible to select the component of the Stokes parameter of which one wants to measure the relative intensity. Then, using

    \[S_c=\frac{I_{\sigma_+} - I_{\sigma_-}} {I_{\sigma_+} + I_{\sigma_-}}
\]

with the measured intensities $I_{\sigma_+}$ and $I_{\sigma_-}$ of the two circularly polarized components, we calculate the circular component of the Stokes vector (Fig.~4(a) of the manuscript).

Fig.~\ref{fig:s2} shows a sketch of the linear wave dynamics in the \textit{x-y}-plane of the microcavity. The polaritons propagate along the positive direction of the \textit{y} axis and are scattered by a defect giving rise phase singularities at points where the amplitude vanishes, i.e. at the dark notches of the intensity profile. The total detected polariton field is given by the interference of the incoming wave and the scattered wave.

\begin{figure}[h]
\includegraphics[scale=0.5]{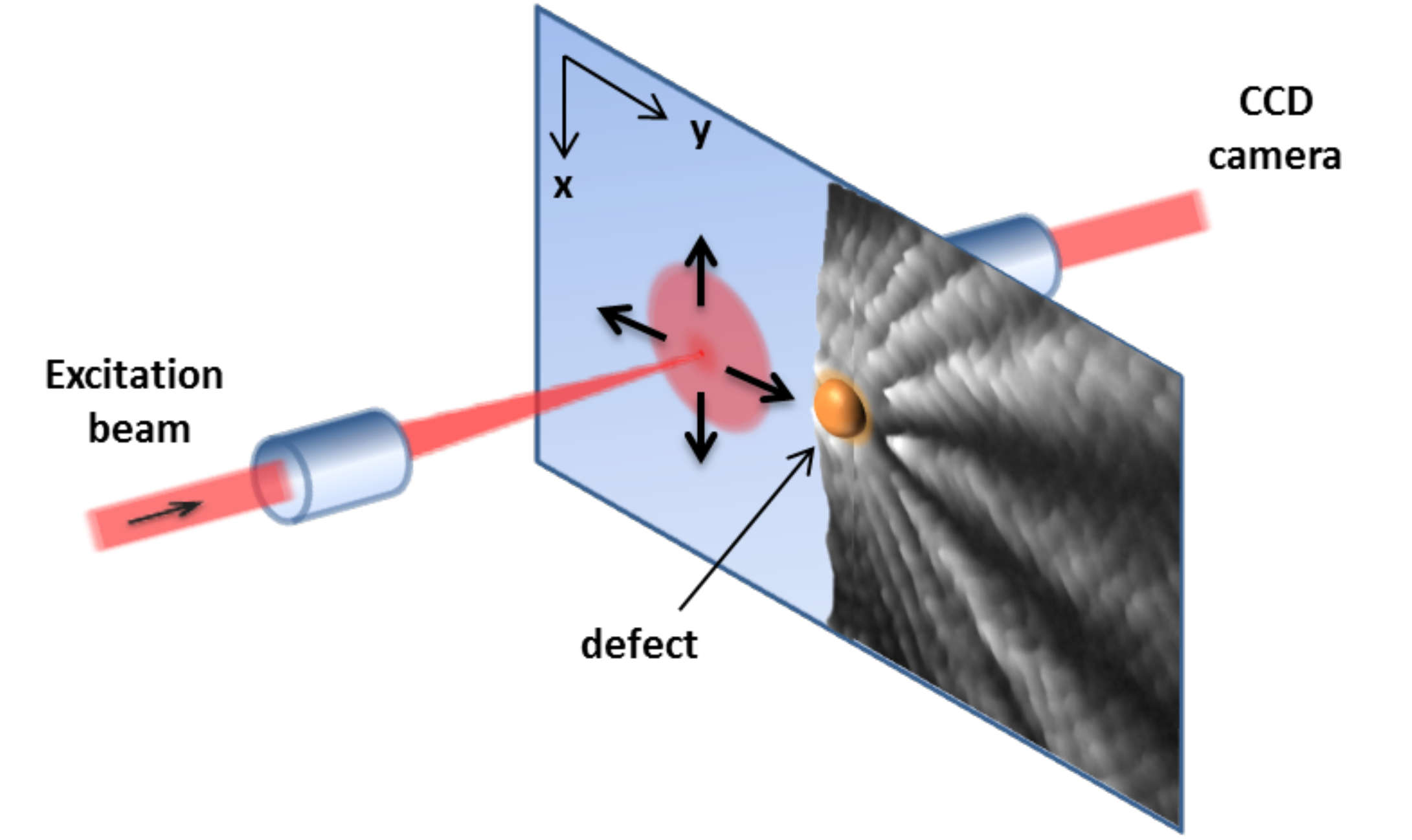}
\caption{Sketch of the linear wave dynamics in the \textit{x-y}-plane of the microcavity.}
\label{fig:s2}
\end{figure}

\newpage
\section{Power dependence measurements and calculation of the polariton density and renormalization}
\label{sec:2}

In this section we demonstrate that the two ``dark-soliton fingerprints'' do not depend qualitatively on the excitation power or on the shape of the excitation beam. 
\\
\\
To verify the linear regime, we studied the excitation density dependence of our results by performing power dependent measurements with both Gaussian and half-Gaussian excitation beams.

In Fig.~\ref{fig:s3} we show the data acquired with a half-Gaussian excitation beam, focused to a spot with FWHM of $\unit[3.5]{\mu m}$ (see Fig.~\ref{fig:s3}e), $\unit[90]{\mu m}$ away from the defect. Here we vary the excitation power by about 5 orders of magnitude, from $\unit[20] {mW}$ (Figs.~\ref{fig:s3} a,b) to $\unit[400] {nW}$ (Figs.~\ref{fig:s3} c,d) and do not observe significant changes in the spatial structure. Two  notches in intensity together with phase vortex lines in the interferograms appear in the linear regime as the result of the scattering and interference. This is confirmed by the intensity profile and by the FWHM calculated for the left dark notches in Figs.~\ref{fig:s3}(a) and \ref{fig:s3}(c), which is respectively $4.2~\pm~0.5$ at $\unit[20]{mW}$ and $3.7~\pm~1$ at $\unit[400]{nW}$ (see Fig.~\ref{fig:s3} f).
\begin{figure}[h]
\includegraphics[scale=0.5]{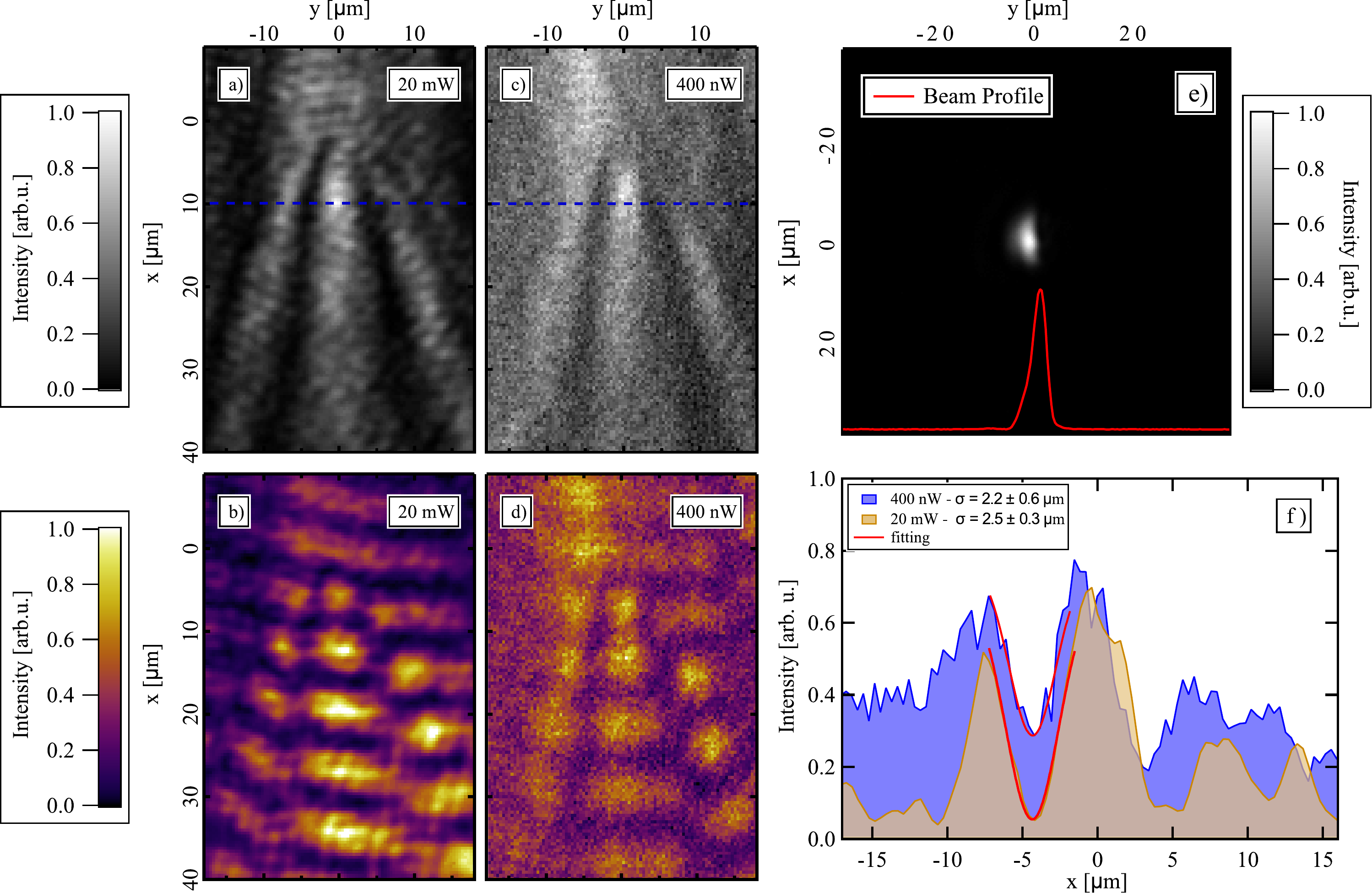}
\caption{Measured real space intensity (a), (c) and interference patterns (b), (d) acquired at $\unit[20]{mW}$ and $\unit[400]{nW}$ excitation power. (e) Half-Gaussian excitation spot. (f) Intensity profiles calculated along the dashed blue lines in a) and c), $10~\mu m$ away from the defect. The red line represents the inverse Gaussian fitting used to calculate the FWHM  of the left dark notches in Figs. (a) and (c). The $\sigma$ of the two fits used for the calculation of the FWHM $2\sqrt{\ln2}\sigma$ are also reported in the legend. The $\unit[400]{nW}$ profile (blue) reported in Fig. (f) has been shifted to the left by about $\unit[1]{\mu m}$ for clarity.}
\label{fig:s3}
\end{figure}

The half-Gaussian beam was created by a confocal excitation scheme as described in the supplementary information of Ref.~\cite{amo_polariton_2011}.  In this scheme, a razor blade is placed between two lenses (at a distance equal to the focal lengths of the lenses) before the excitation objective (Obj$_1$ in Fig.~\ref{fig:s1}). The first lens focuses the laser on a razor blade while the second collimates the image of the razor blade on the sample. In this way, by adjusting the position of the razor blade it is possible to shape the excitation beam to a half-Gaussian profile (Fig.~\ref{fig:s3} e).
Fig.~\ref{fig:s3} shows the data acquired with this scheme. Also in this case, as in the case of a Gaussian excitation beam (Figs.~1,~2,~3,~4 of the main manuscript and Figs.~\ref{fig:s10},~\ref{fig:s7},~\ref{fig:s8} of the supplementary information), we observe the formation of the two ``dark-soliton fingerprints''.
\\
\\
 In the microcavity polariton  literature, the observation of dark solitons has been claimed using both a Gaussian excitation beam~\cite{amo_polariton_2011, PhysRevB.86.020509} and half-Gaussian excitation beam~\cite{amo_polariton_2011, hivet_half-solitons_2012}. We have used both excitation shapes and find that they do not affect the observed structure significantly.
\\
\\
\textbf{Polariton density}. The polariton density inside the cavity has been estimated from the number of photons transmitted through the sample and detected on the CCD camera. For a microcavity, in fact, the polariton population is proportional to the detected intensity. At the lowest excitation density of $\unit[3.8]{W/cm^{2}}$, the polariton density inside the microcavity is estimated to be $D_{pol}=\unit[2.3\times 10^2]{cm^{-2}}$, seven orders of magnitude lower than the lasing threshold observed in standard microcavities~\cite{Deng23122003}. The density of polaritons have been estimated by using the following formula:
\begin{equation}
D_{pol}= \Phi_{hv}\times\tau,
\end{equation}
where $\Phi_{hv}= \unit[2.3\times10^{13}]{cm^{-2}~s^{-1}}$ is the flux of photons transmitted through the sample and $\tau=10~ps$ is the polariton lifetime.
The flux of photons $\Phi_{hv}$ has been calculated as:
\begin{equation}
\Phi_{hv}= \frac{C_{px} \times \alpha_{phe}}{t\times Q \times\ A_{px} \times \eta_{obj} \times \eta_{lens}},
\end{equation}
where $C_{px}=7186$ is the maximum pixel counts corrected for the background, $\alpha_{phe} = 7.3$ is the number of photoelectrons per count (determined from the shot-noise), $t=10\,$s is the integration time, $Q=0.3$ is the quantum efficiency of the CCD camera at the wavelength used in the experiment and $A_{px}= 0.1225~\unit{\mu m^2}$ is the real-space area of a single CCD pixel on the sample, $\eta_{obj}=0.7$ and $\eta_{lens}=0.9$ are the assumed intensity transmission factors due respectively to the objective  ($Obj_2$) and the lens ($L_2$) used in the experiment (see Fig.~\ref{fig:s1}).

The investigated sample has a large negative detuning of -29\,meV of the cavity photon mode from the exciton resonance (1.514\,eV), resulting in an exciton fraction of the polariton mode of less than 1\%. The interaction energy scales with the excitonic content, and can be estimated using Eq.~(9) in~\cite{PhysRevB.70.205301}. For the highest density used ($1.1\times 10^8\,$cm$^{-2}$), we find a renormalization energy of 33\,neV, using the bulk exciton binding energy of 4.2\,meV, a Rabi splitting of 5\,meV, the bulk GaAs exciton Bohr radius of 14\,nm, and an excitonic fraction of 1\%. The parameters used here are approximate values, and the resulting renormalization is an order of magnitude estimate. However, since it is three orders of magnitude lower than the polariton linewidth, it is sufficiently accurate to predict that it has a negligible effect on the polariton dynamics, consistent with the experimental observation.

\newpage
\section{Energy dependent measurements for a different defect}
\label{sec:3}
In this section we demonstrate that the disappearance of the two ``dark-soliton fingerprints'' with detuning can be observed for different defects and can be considered a typical behaviour.
\begin{figure}[!hbtp]
\includegraphics[scale=0.55]{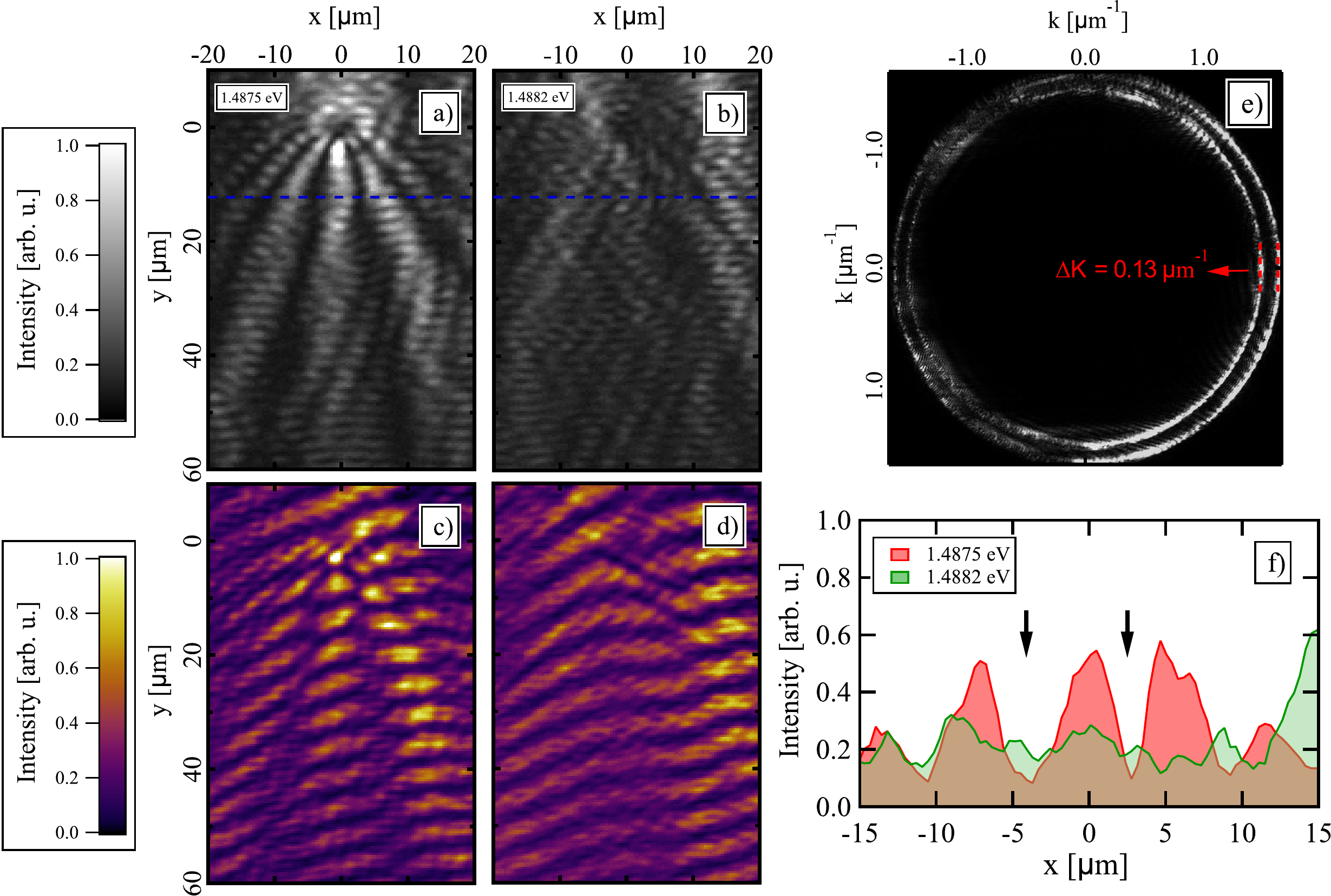}
\caption{Experimental real space intensity (a), (b) and interference patterns (c), (d) showing the appearance (at $\unit[1.4875] {eV}$) and disappearance (at $\unit[1.4882]{eV}$) of both dark notches and phase shifts when the energy of the excitation beam is increased. In this case the excitation spot is focused $\unit[50]{\mu m}$ away from the defect. (e) Experimental K-space images showing the increase of polariton wave vector with the increase of the energy. The measured increases is $\unit[0.13]{\mu m^{-1}}$. (f) Intensity profiles calculated along the blue dashed line, $\unit[12]{\mu m}$ downstream of the defect confirm the disappearance of the features (green profile). The two arrows indicate the positions of the solitonlike features.}
\label{fig:s10}
\end{figure}

By using the same experimental setup described in section~\ref{sec:1}, we study the scattering of polaritons from various defects and observe the appearance and disappearance of the solitonlike features when the energy of the excitation beam and consequently the wave vector of polaritons is varied. 
Depending on the defect, different variation of the polariton wave vector is needed in order to observe the transition. The increase/decrease of the polariton wave vector changes the interference condition of the polaritons scattered by the defect so that both the dark notches and the phase dislocations disappear. In the case of Fig.~\ref{fig:s10}, the dark-solitonlike features disappear when the energy of the the excitation beam is increased by $0.7$ meV, which corresponds to an increase of the wave vector of $\unit[0.13]{\mu m^{-1}}$, as shown in Fig.~\ref{fig:s10}(e). The disappearance of the dark notches is confirmed by the intensity profiles extracted $\unit[12]{\mu m}$ away from the defect (Fig.~\ref{fig:s10} f). We have investigated several defects and found that a similar transition could be observed by decreasing the wave vector of the propagating polariton mode since also in this case the interference condition changes.

\section{Evaluation of the dark soliton conditions for measurements in the linear regime}
\label{sec:3}

Since the first observation of dark solitons in atomic Bose-Einstein condensates (BECs)~\cite{denschlag_generating_2000}, the soliton speed ($v_s$) has been expressed in terms of either the phase shift ($0 < \delta < \pi$) or the soliton depth ($n_s$), the latter being the difference between the density of the condensate ($n$) and the density at the bottom of the notches ($n_{d}$):

\begin{equation}\label{eq:solve}
\frac{v_{s}}{c_{s}} = \cos~\frac{\delta}{2} = \sqrt{ {1-} {\frac{n_s}{n}}} 
\end{equation}

In the above formula, $c_s$ represents the speed of sound in the BEC, which is directly proportional to the density of the condensate~\cite{denschlag_generating_2000}. Equation~\ref{eq:solve} dictates the conditions that have to be satisfied in order to identify dark solitons in BECs. In particular, when $\delta$ tends to $\pi$: 

\begin{enumerate}
	\item ${\frac {n_s} {n}} \rightarrow 1 $, i.e. the dark soliton becomes deeper ($n_s = n_{d}$)
	\item ${\frac {v_s} {c_s}} \rightarrow 0 $, i.e. the velocity of the dark solitons tends to zero.
\end{enumerate}

Furthermore, due to the repulsive interactions within the BEC, the width of the dark soliton tends to the healing length $\xi$ of the condensate~\cite{denschlag_generating_2000, PhysRevA.58.2417}:
\begin{center}
$\xi \rightarrow \left({\frac{2nMg}{\hbar^2} }\right)^{-1/2}$. 
\end{center}
Here \textit{M} is the atomic mass, \textit{g} is the atom-atom interaction and $\hbar$ is the Planck constant divided by $2\pi$.

As in the atomic BECs, also in the case of polariton condensates equation~\ref{eq:solve} has been used to identify dark solitons~\cite{amo_polariton_2011,PhysRevLett.107.245301, PhysRevB.86.020509, deveau_book}. 
However, in all the previous works the healing length condition which is related to the width of the soliton has been neglected. This aspect will be discussed in the last part of this section. 

First we evaluate the conditions 1 and 2 in the linear regime. 
\\
\\
\textbf{Depth and velocity of the dark notch.} In our case, as for the polariton quantum fluid, dark notches are characterized by a minimum of the intensity ($n_{d}$) at their center compared to the surrounding polaritons ($n$). Consequently, we can evaluate the depth ($n_s$) of a dark notch in the same way as in the quantum fluid case:
\begin{equation}
n_s = n - n_d
\end{equation}
and calculate the relative depth of the dark notch (${\frac {n_s} {n}}$) at different distances from the defect, where the phase shift $\delta$ is close to $\pi$. 

Fig.~\ref{fig:s4}(c) shows the relative depth (${\frac {n_s} {n}}$) of the dark notch for the left notch of Fig.~\ref{fig:s4}(a). The depth of the dark notch has been determined by fitting the line profile of the notch with an inverse Gaussian distribution at different distances from the defect. The intensity of the surrounding polaritons ($n$) has been estimated from the maximum intensity along the red dotted line in Fig.~\ref{fig:s4}(a). The fitting (red line) and the quantities $n$, $n_s$ and $n_d$ are shown in Fig.~\ref{fig:s4}(c).
\begin{figure}[!hbtp]
\includegraphics[scale=0.7]{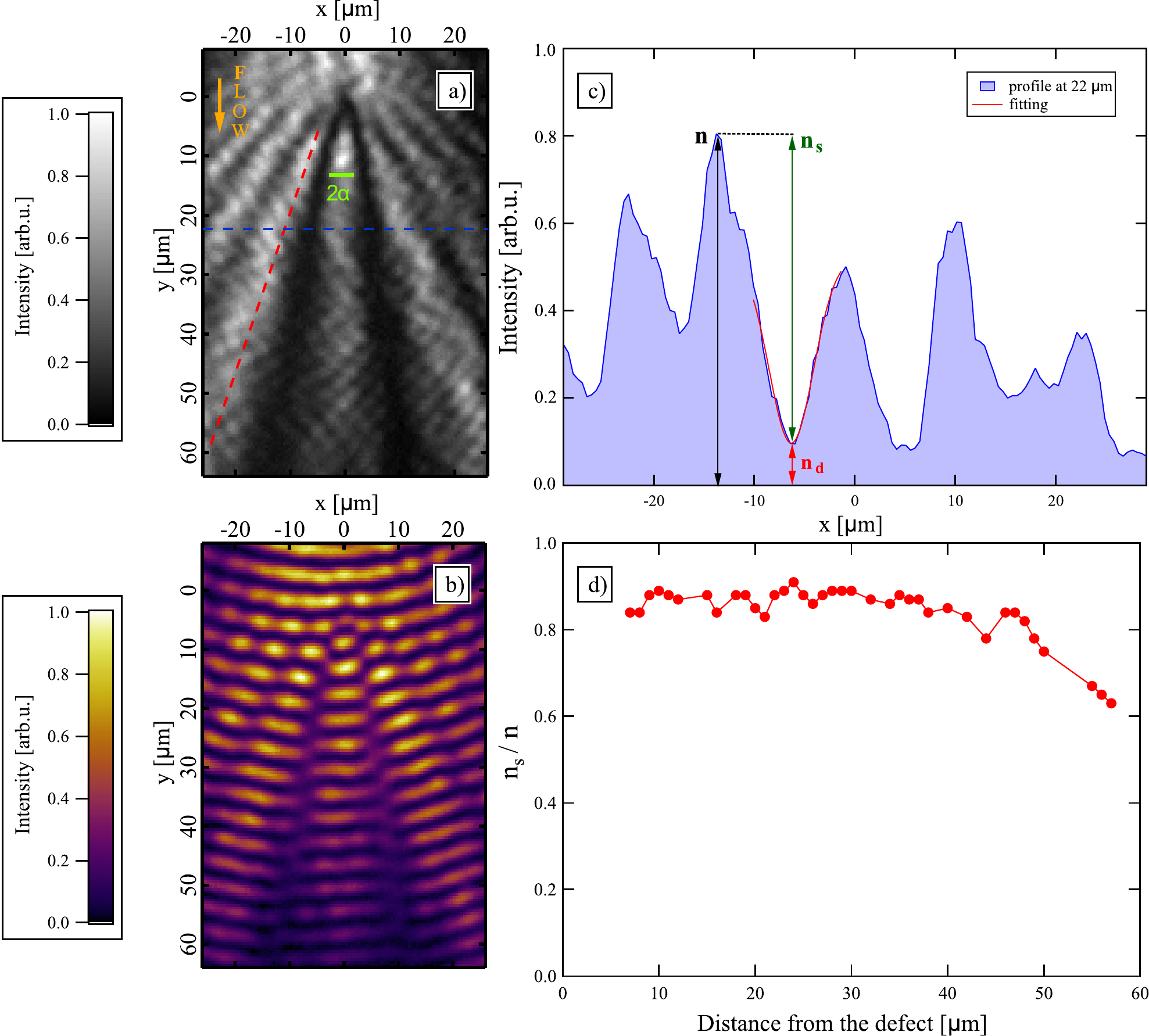}
\caption{ Experimental real space intensity (a) and interference (b) patterns showing the two ``dark-soliton fingerprints''. These images are the same as Figs.~1(a) and (b) of the main manuscript but plotted over a larger \textit{y} axis range. (c) Horizontal intensity profile calculated along the blue dashed line in (a), $\unit[22] {\mu m}$ away from the defect. The inverse Gaussian fit is also shown (red line) together with the quantities $n$ (black arrow), $n_s$ (green arrow) and $n_d$ (red arrow). (d) Dark-notch depth (${\frac {n_s} {n}}$) calculated from (a) at different distances from the defect. As in the polariton quantum fluid case, the depth of the dark notch is stable up to $\unit[42]{\mu m}$.  }
\label{fig:s4}
\end{figure}

Similar to the case of a quantum fluid, in our linear system the relative depth of the dark notch remains approximately constant up to $42~\unit{\mu m}$ as shown in Fig.~\ref{fig:s4}(d), which corresponds to a close to $\pi$ phase shift in the interferogram (Fig.~\ref{fig:s4} b). It is worth noticing that the ratio $n_s/n$ oscillates around the mean value of $0.87$ and it reaches the max value of $0.91$ at $24~\mu$m away from the defect, similar to the critical value of $0.9$ reported in the literature~\cite{PhysRevLett.107.245301,deveau_book} for the formation of ``vortex streets''. 

Moreover, at longer distances, at about $50~\unit{\mu m}$ away from the defect we observe a decrease of the relative depth of the dark notch together with a decrease of the phase shift. In agreement with equation~\ref{eq:solve}, when the dark notch is deeper ($n_s/n$ tends to one in the equation~\ref{eq:solve}) the ratio $v_s/c_s$ decreases.
\\
\\
These observations show that a similar trend as in the polariton quantum fluid can be observed also in the linear regime. 
\\
\\
\textbf{Equivalent of the Mach number.} In the case of a polariton quantum fluid it has been shown that different hydrodynamic regimes are connected to the Mach number (\textit{M}), which is the ratio between the local flow velocity $v_{flow}$ and the local speed of sound $c_s$~\cite{amo_polariton_2011,PhysRevLett.107.245301, PhysRevB.86.020509, deveau_book}:

\begin{equation}\label{eq:mach}
M = \frac{v_{flow}} {c_s}.
\end{equation}

Dark solitons in polariton microcavities have been claimed to appear for $M>1$, with values depending on the nature of the obstacle~\cite{amo_polariton_2011,PhysRevLett.107.245301, PhysRevB.86.020509, deveau_book}.
In our linear system there is no sound, i.e. no linear dispersion range. However, we can evaluate the equivalent of the Mach number in our system, namely taking ($\frac{v_{flow}} {c_s}$) from the measured values of $n_s/n$ (Fig.~\ref{fig:s4} d) and using the geometrical relation~\cite{amo_polariton_2011,PhysRevLett.107.245301, PhysRevB.86.020509, deveau_book}: 

\begin{equation}\label{eq:angle}
v_{s} = v_{flow}~\sin(\alpha)  
\end{equation}
where $v_{flow}$ is the velocity of the polariton flow along the \textit{y}-direction and $\alpha$ is the aperture angle of the oblique-dark notch with respect to the flow direction. In our case $\alpha$ is measured to be $16.3^{\circ}$.

By combining equations~\ref{eq:solve},~\ref{eq:mach} and~\ref{eq:angle} we can estimate the equivalent \textit{M} of the quantum fluid case:
\begin{equation}\label{eq:mym}
M = \sqrt{1-\frac{n_s}{n}}~\frac{1}{sin(\alpha)} 
\end{equation}

In our case the ratio $\frac{n_s}{n}$ varies between $0.91$ and $0.63$ (measured respectively at $24~\unit{\mu m}$ and $57~\unit{\mu m}$ from the defect (see Fig.~\ref{fig:s4}(d), which corresponds to a variation of \textit{M} from a minimum value of $1.07$ to a maximum value of $2.17$. Dark solitons have been predicted to appear in polariton microcavities when $M\geq1.02$~\cite{pigeon_hydrodynamic_2011}. 
This analysis shows that the condition on the Mach number to observe dark solitons can be satisfied along the whole path of the dark notch in the linear regime. 
\\
\\
Therefore, $M>1$ together with the constant relative depth of the dark notch ($\frac{n_s}{n}$) and related phase shifts in the interferograms are conditions necessary but not sufficient to identify dark solitons, since these conditions can be observed also in the linear regime.
\\
\\
\textbf{Healing length condition.} Equation~\ref{eq:solve}, initially proposed for atomic BECs~\cite{denschlag_generating_2000}, has been used also in polariton microcavities to identify dark solitons~\cite{amo_polariton_2011,PhysRevLett.107.245301, PhysRevB.86.020509, deveau_book}. However, in all the previous works, the healing length condition which is related to the width of the soliton has been neglected even though it represents a direct application of the definition of a soliton. 
\\
A soliton, in fact, is a solitary wave that preserves its shape while propagating through a dispersive medium~\cite{stegeman_optical_1999,segev_self-trapping_1998}. This feature can be considered as universal fingerprints since has been observed in all the physical systems where solitons have been studied~\cite{Segev:02}. To the best of our knowledge the formation of ``oblique" dark solitons, although predicted for both the atomic~\cite{PhysRevLett.97.180405} and polariton condensates~\cite{pigeon_hydrodynamic_2011}, has been experimentally reported only for polariton condensates~\cite{amo_polariton_2011,PhysRevLett.107.245301, PhysRevB.86.020509, deveau_book}. On the other hand, in atomic BECs, only single ``straight" dark solitons have been experimentally observed~\cite{denschlag_generating_2000}.

The above healing length condition specifies that solitons propagating in a condensate of homogeneous density are characterized by a constant width (i.e. nonspreading wave) which is given by the healing length $\xi$ of the condensate~\cite{denschlag_generating_2000}. When the excitation density is increased, the FWHM of the dark notch should scale as $n^{-1/2}$, proportional to the healing length of the condensate. 

We compare in Fig.~\ref{fig:s5} the measured width of the dark notch in our linear system with the expected scaling $C/n^{1/2}$ using the measured intensity as function of distance. We find that the healing length condition is not respected by this data in the linear regime, indicating that it is suited to discriminate dark solitons from linear propagation. 
We therefore propose to use the healing length condition to verify dark soliton formation, which should be fulfilled over a range of polariton excitation densities to exclude coincidental matches with specific scattering patterns in linear propagation.  
\begin{figure}[!hbtp]
\includegraphics[scale=0.6]{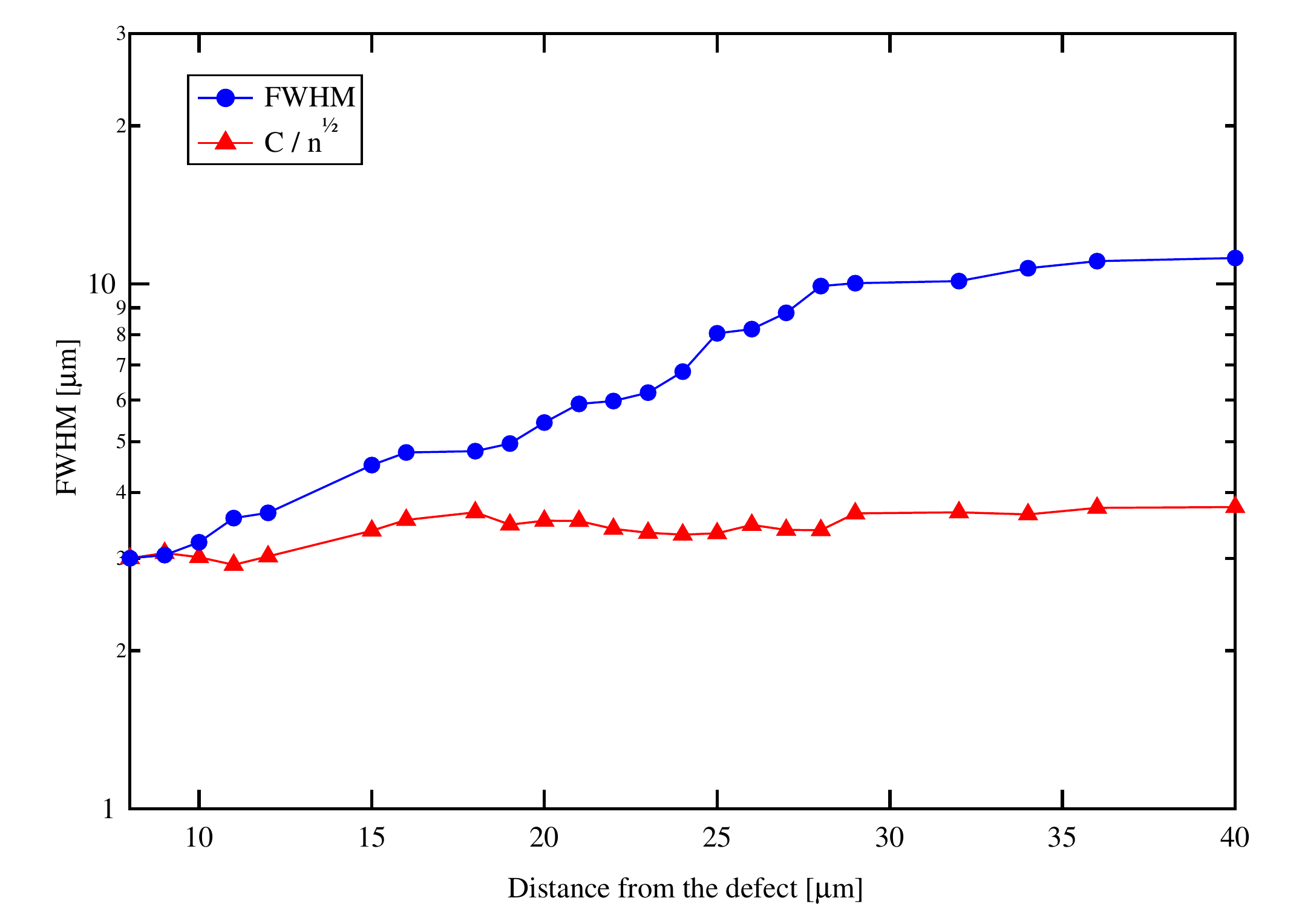}
\caption{The blue circles show the measured FWHM of the left notch in Fig.~\ref{fig:s4}(a) at different distances from the defect. The red triangles show $C/n^{1/2}$ proportional to the healing length using a suitable chosen constant $C$. The density $n$ has been calculated by averaging the intensity of the left and right sides of the left dark notch in Fig.~\ref{fig:s4}(a).}
\label{fig:s5}
\end{figure}

\newpage
\section{Theory for the cavity mode scattering by a circular defect}
\label{sec:4}

The classical theory of electromagnetism is used in order to calculate the distribution of electric and magnetic fields inside the cavity
in the presence of a disk-shaped defect and illumination of the cavity by a monochromatic laser beam. As mentioned in the manuscript, the choice of such a model is justified by the fact that we operate in the linear regime with a polariton dispersion dominated by the cavity mode. 
In the model, we consider the propagation of two-dimensional photons with a quadratic dispersion in the microcavity plane, as shown in Fig.~\ref{fig:s6}. A quadratic dispersion is found for all planar microcavity polaritons for small in-plane momenta plane. In our case, the large negative detuning provides a large range over which the dispersion is to a good approximation quadratic, covering all the relevant excitation wave vectors used.

The field distribution in a bare cavity obeys Maxwell's equations for the electric field $\vect{E}(x,y,z,t)$ and  magnetic field
$\vect{H}(x,y,z,t)$. Symmetry of the planar cavity allows one to separate the solutions as follows:
\begin{eqnarray*}
  \vect{E}(x,y,z,t)&=&\vect{E}_\omega(x,y)\chi(z)\exp[{-}i\omega t]\\
  \vect{H}(x,y,z,t)&=&\vect{H}_\omega(x,y)\xi(z)\exp[{-}i\omega t]
\end{eqnarray*}
The subscript $\omega$ denotes that the in-plane components of the fields depend on the energy of radiation while the normal components $\chi(z)$ and $\xi(z)$
are independent of energy under consideration of small in-plane wave vector $k_\parallel$:
$$
  k_\parallel\ll\frac{n_{\rm cav}\omega}{c}
$$
where $n_{\rm cav}$ is the refractive index of material of the cavity and $c$ is the vacuum speed of light. The normal components $\chi(z)$ and $\xi(z)$ of
the fields can be estimated inside the cavity where the most of light energy is concentrated: $\chi(z)\propto\xi(z) \propto\cos( n_{\rm cav}\omega_0 z/c)$ where $\omega_0$ is the cavity resonance frequency at normal incidence.
The in-plane wave vector then can be deduced as
$$
k_\parallel(\omega)=\frac{n_{\rm cav}}{c}\sqrt{\omega^2-\omega_0^2}.
$$
The cavity defect is given by a change of the Bragg mirror composition by the presence of additional GaAs due to the Ga droplet formation during the growth process~\cite{fujiwara_classification_1987, chand_comprehensive_1990}. The presence of the defect has the effect to modify the effective thickness of the cavity layer, resulting for example in a red-shift of the photonic dispersion inside the defect~\cite{zajac_structure_2012}.
As a result, the resonance frequency inside the defect shifts from $\omega_0$ to $\omega_0'$ with respect to the bare cavity and accordingly the in-plane wave vector $k_\parallel$ to $k_\parallel'$ , with $k_\parallel' > k_\parallel\ $ (Fig.~\ref{fig:s6}). The energy shift of the cavity mode represents an attractive potential in the two-dimensional polariton propagation.

\begin{figure}[h]
\includegraphics[scale=0.7]{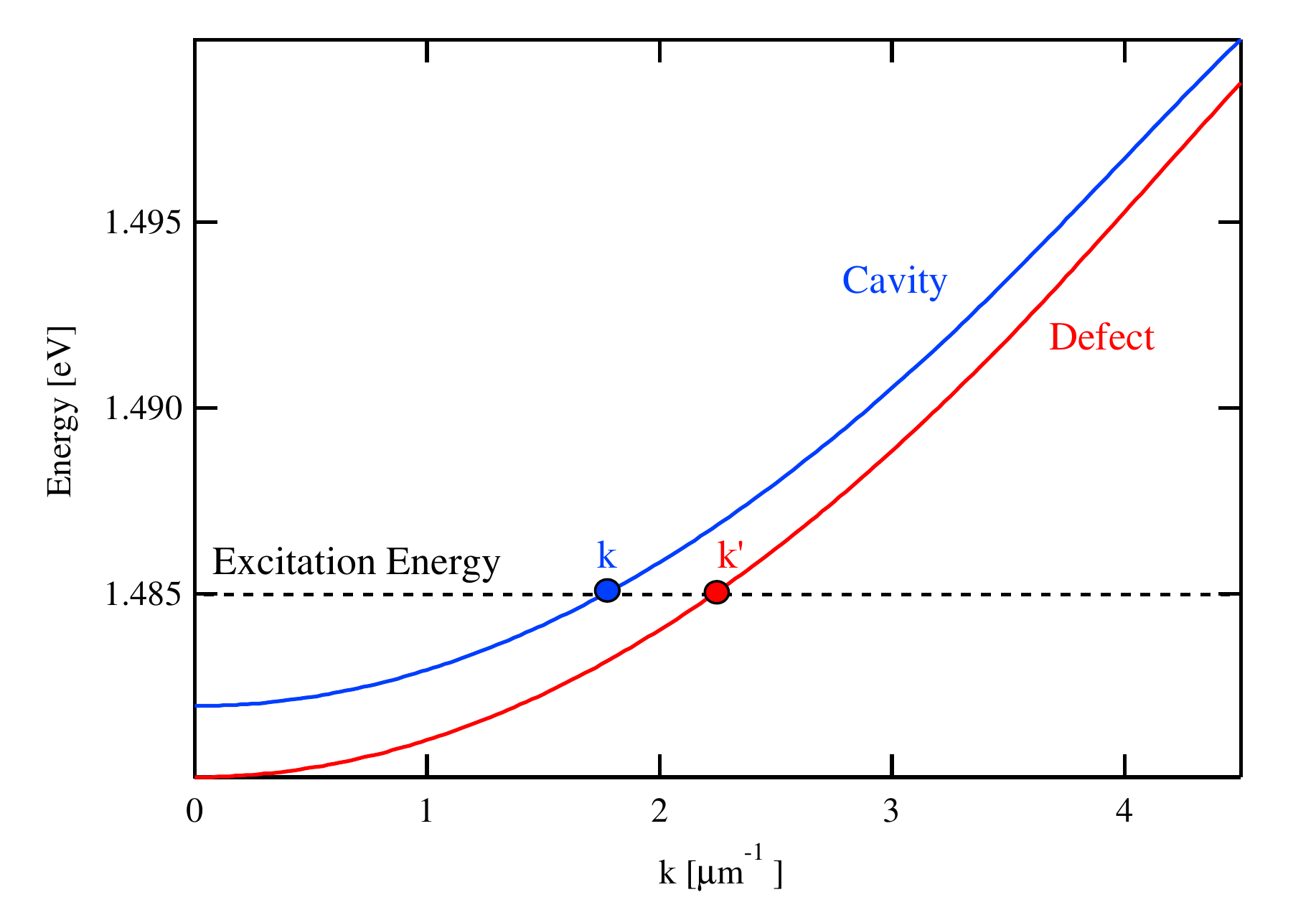}
\caption{Theoretical dispersion inside (red) and outside (blue) the defect. The presence of the defect has the effect to modify the effective thickness of the cavity layer, resulting in a red-shift of the photonic dispersion inside the defect. Consequently, for a fixed energy, the wave vector of the photonic mode in the region of the defect is higher than in the rest of the cavity ($k' > k$). The black dashed line indicate the excitation energy used in the experiment.}
\label{fig:s6}
\end{figure}

Besides the change of the resonance condition, also the normal components of the fields vary the spatial distribution and become $\chi(z)\to\chi'(z)$ and $\xi(z)\to\xi'(z)$. In our model, however, we assume that these changes are small (the relative change of the cavity energy considered in our case is only about 0.1\%) and therefore we neglect them. Within this approximation, the solution
of the problem of light propagation through a cavity with arbitrarily shaped defect is reduced to the solution solely in the $xy$ plane because
boundary conditions are independent of the position on the axis $z$.  First we find two basis sets of solutions of Maxwell's equations for the bare cavity and the perturbed cavity. We denote these sets as $\vect{E}_{\omega,j,m}^\cc,\,\vect{H}_{\omega,j,m}^\cc$ and $\vect{E}_{\omega,j,m}^\dd,\,\vect{H}_{\omega,j,m}^\dd$ respectively. Here the index $j$ stands for polarization (TE or TM) and $m$ is the discrete index of the mode (angular momentum quantum number around the defect center) in the expansion.

The two respective sets of fields defined above are local solutions of Maxwell's equations outside and inside the defect area. In order to solve the whole problem of scattering, we have to find a solution on the boundary between the bare cavity and the defect where the in-plane wave vector is not continuous. Here we assume that the boundary behaves like an ordinary boundary between two dielectrics, i.e. we require continuous tangent components of all fields. Let us write the fields in the bare cavity and in the defect area in the following form:
\begin{eqnarray}
  \vect{E}^\cc_\omega&=&\vect{E}_{\rm incident}+
    \sum_{j,m}c_{j,m}^\cc\vect{E}_{\omega,j,m}^\cc\\
  \vect{H}^\cc_\omega&=&\vect{H}_{\rm incident}+
    \sum_{j,m}c_{j,m}^\cc\vect{H}_{\omega,j,m}^\cc\\
  \vect{E}^\dd_\omega&=&\sum_{j,m}c_{j,m}^\dd\vect{E}_{\omega,j,m}^\dd\\
  \vect{H}^\dd_\omega&=&\sum_{j,m}c_{j,m}^\dd\vect{H}_{\omega,j,m}^\dd
\end{eqnarray}
The coefficients $c^\cc$ and $c^\dd$ are finally set so that the boundary conditions are fulfilled. If the basis sets are chosen properly, the solution is unambiguous. For the case of a circular defect, it is convenient to use the basis of fields in cylindrical coordinates~\cite{snyder_love} whose boundary conditions reduce to simple algebraic equations for the unknown coefficients. Once the coefficients are known, the spatial field distribution is evaluated using the definitions above, performing the summation on right hand side. To include the TE--TM splitting in cylindrical coordinates, it suffices to discriminate between the in-plane wave vectors
$k_{\parallel,\rm TE}$ and $k_{\parallel,\rm TM}$ and the same inside the defect.

\section{Dependence of the solitonlike features on the scattering geometry}
\label{sec:5}

The observed features depend on the shape and size of the defect and the direction and polarization of the incoming polariton wave relative to the defect. For an elliptical defect, the phase and amplitude of the scattering depend on the direction of the incoming wave. Also the polarization contributes to the anisotropy of the effect because for a given absolute polarization direction a different angle of incidence corresponds to a different polarization relative to the defect.

Fig.~\ref{fig:s7} shows an example of the beam incident on the defect at an angle in the experiment. We use the same parameters of Figs.~1(c) and 1(d) of the manuscript to perform the simulations, but we change the direction of the incoming beam. In the previous case (Fig.~1 of the manuscript), the excitation beam is polarized orthogonal to the incidence direction, while in Fig.~\ref{fig:s7} the beam direction has a 28 degree angle to its polarization ($y$), and generates a phase dislocation only in the upper dark line but not in the lower one, as indicated by the arrow in Fig.~\ref{fig:s7}(c).
\begin{figure}[h]
\includegraphics[scale=0.5]{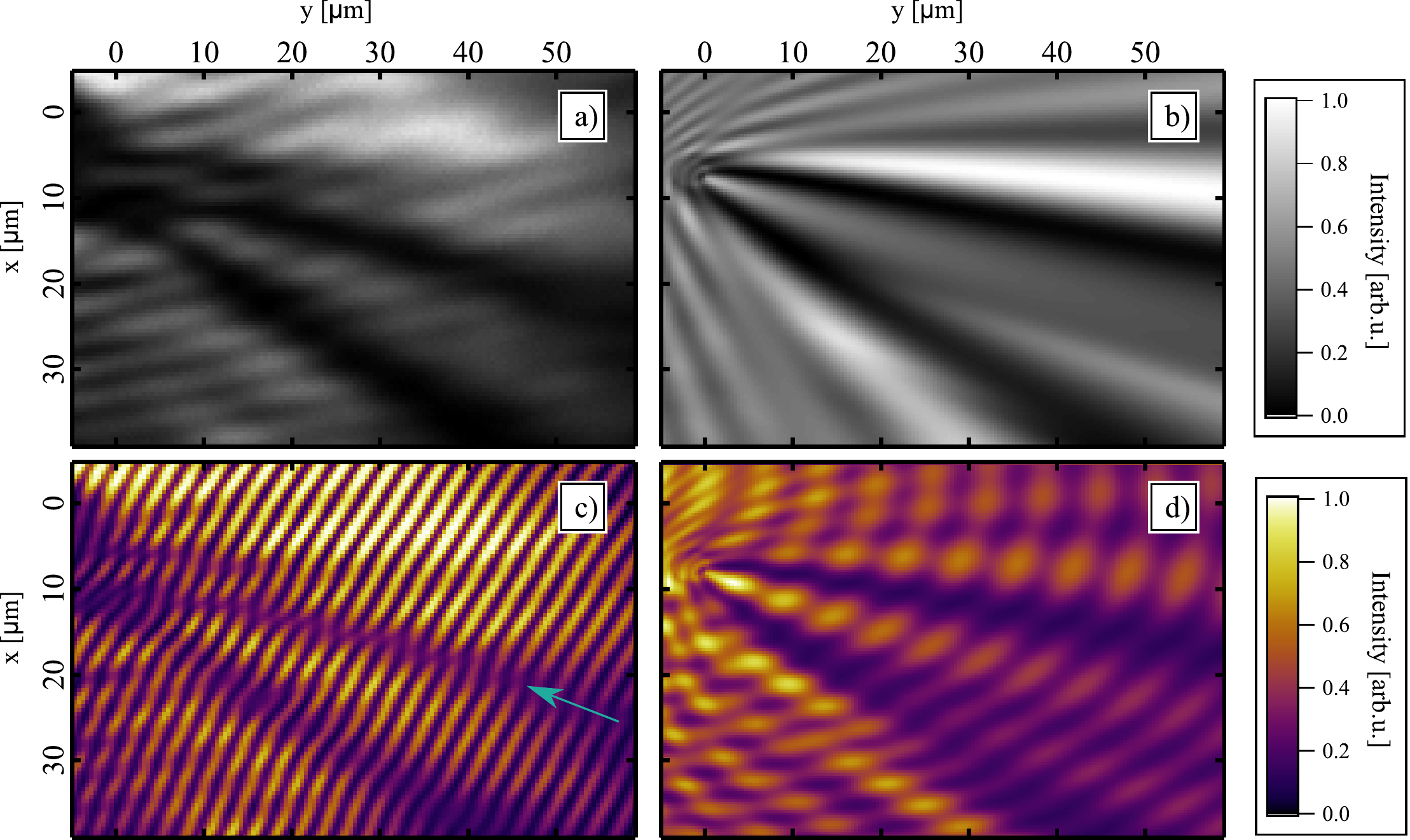}
\caption{Experimental (a),(c) and simulated (b),(d) real-space intensity and interference pattern showing dark solitons fingerprints generated by the interaction of the beam with a defect. Unlike Fig.~1 of the manuscript, the phase shift is only present in correspondence of the upper solitonlike feature as indicated by the light blue arrow in (c). }
\label{fig:s7}
\end{figure}
\\
\\
Moreover, we have investigated the case of a larger defect. The number of dark lines increases with increasing defect size, allowing the formation of quadruplet solitons-like features. This is confirmed by the simulations shown in Figs.~\ref{fig:s8}(b) and \ref{fig:s8}(d). Once again we refer to the simulations shown in Fig.~1 of the manuscript to simulate high-order dislocations. In particular, Figs.~\ref{fig:s8}(b) and \ref{fig:s8}(d) have been calculated by using the same parameters as Figs.~1(c) and 1(d) of the manuscript except for increasing the radius of the defect from 3\,$\mu$m to 5\,$\mu$m.

In Figs.~\ref{fig:s8}(a) and \ref{fig:s8}(c) the experimental observation of a high order solitonlike features is shown in both intensity and phase. In the case of a bigger defect, it is possible to note how the wave appears to bend around the edges of the defect.
\begin{figure}[h]
\includegraphics[scale=0.5]{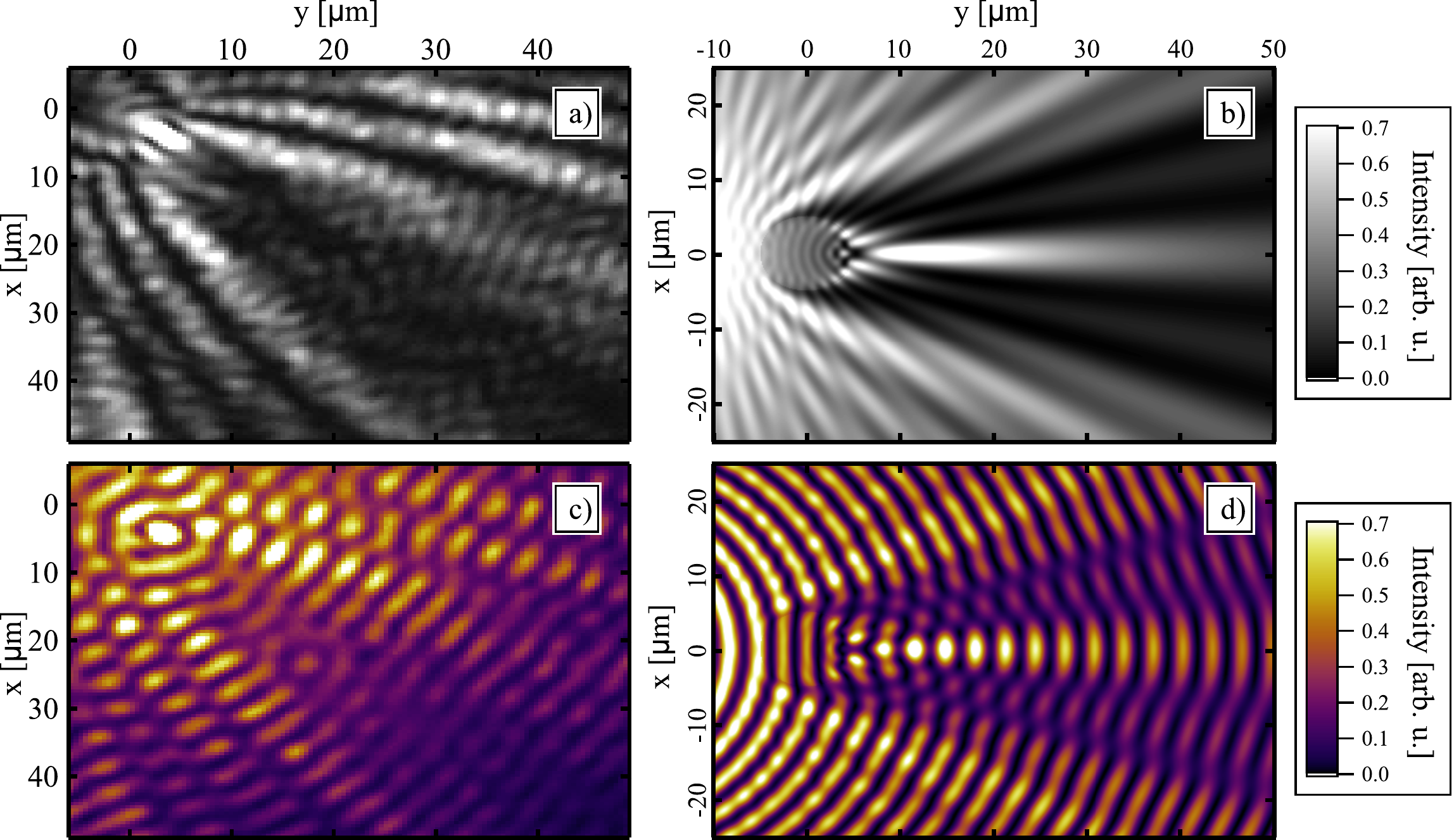}
\caption{Experimental (a),(c) and simulated (b),(d) real-space intensity and interference pattern showing higher-order soliton features generated by the interaction of the beam with a defect bigger than the one present in Fig.~1 of the manuscript. }
\label{fig:s8}
\end{figure}

\newpage
\section{Half-solitonlike features caused by TE-TM splitting}
\label{sec:6}

In our simulations a linear \textit{y}-polarized incoming beam, propagates along the \textit{y} direction and is scattered by a defect positioned at $\unit[25]{\mu m}$ away from the excitation spot, inducing the formation of two traces propagating in oblique directions.
In the case of half-soliton features in the circular polarisation basis, we found that the birefringence in the scattering by the defect can be explained by the intrinsic TE-TM splitting of the polariton dispersion. This is confirmed by the simulations shown in Fig.~\ref{fig:s9} where the scattered field, produced by the wave hitting the defect, is calculated in absence, Fig.~\ref{fig:s9}(a), or in presence, Fig.~\ref{fig:s9}(b) of the TE-TM splitting. In the latter case we use $\vec{k}_{\parallel L}$/$\vec{k}_{\parallel T}=1.004$ which is the same value that has been used in reference~\cite{maragkou_optical_2011} for the same sample. In order to simplify the theoretical discussion, we consider the TE-TM splitting constant across the whole cavity including the defect and no additional splitting in the defect is considered.
\begin{figure}[!hbtp]
\includegraphics[scale=0.5]{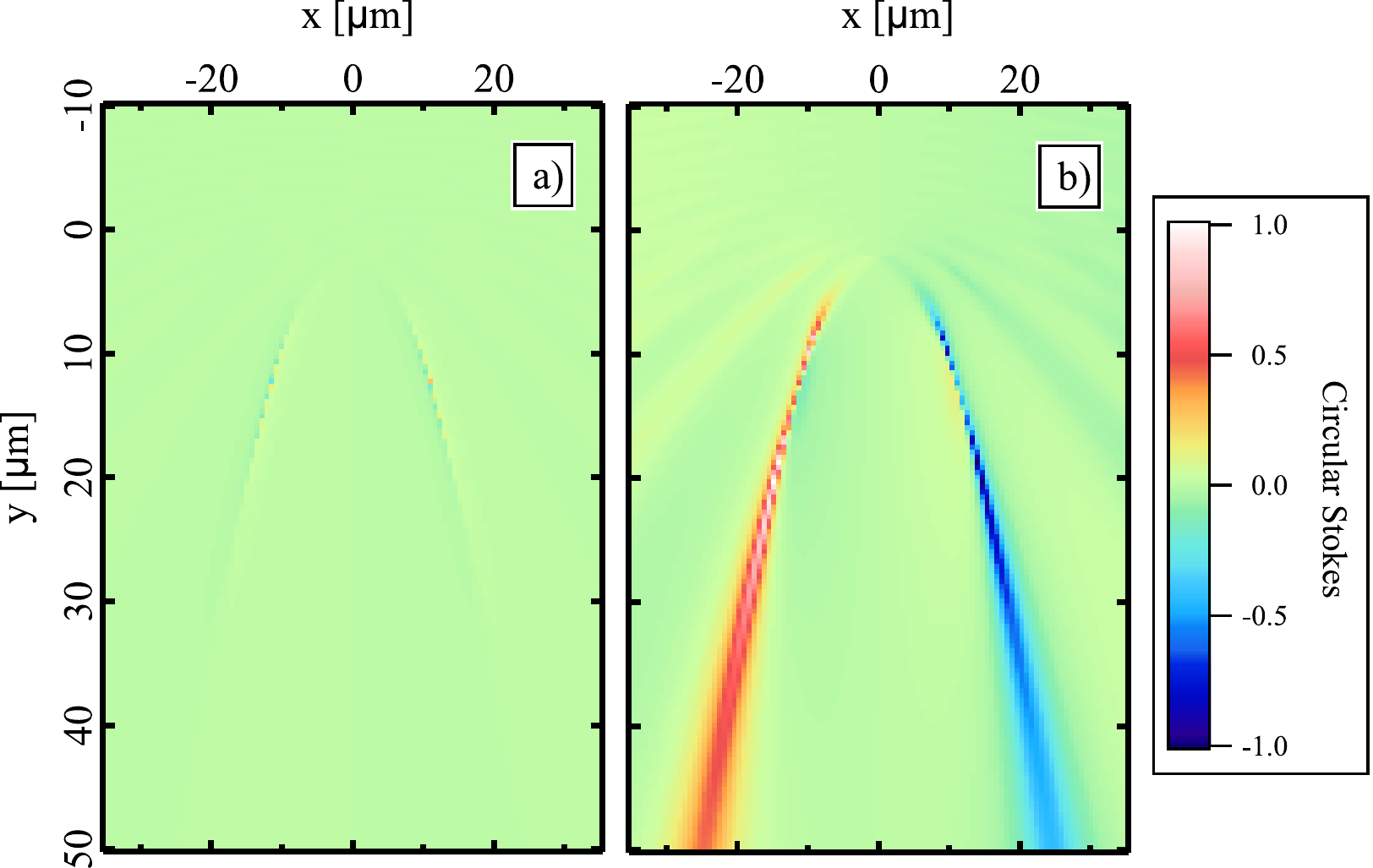}
\caption{Simulated circular Stokes parameters showing half-soliton features. The images have been calculated by considering a beam hitting a circular defect in absence (a) and in presence (b) of the TE-TM  splitting.}
\label{fig:s9}
\end{figure}

\end{widetext}

\newpage
\bibliography{Bibliography}
\end{document}